\documentclass[
     ,final            
   ]
   {aipproc}

\layoutstyle{6x9}

\begin{document}

\title{A Unified Model for inelastic $e-N$ and $\nu-N$ cross sections
at all $Q^2$}
\classification{12.38.-t,12.38.Mh,12.38.Qk,13.60.Hb}
\keywords      {structure functions, unified model}

\author{Arie Bodek}{
  address={Department of Physics and Astronomy,  University of Rochester,
Rochester, New York 14618,  USA}
}
\author{Un-ki Yang}{
  address={Enrico Fermi Institute, University of Chicago,Chicago,
  Illinois 60637, USA}
}

\begin{abstract}
We present results using a new scaling variable, $\xi_w$
in modeling electron- and neutrino-nucleon scattering cross sections
with effective leading order PDFs.
Our model uses all inelastic charged lepton
$F_2$ data (SLAC/BCDMS/NMC/HERA), and photoproduction data
on hydrogen and deuterium. We find that our model describes all inelastic
scattering charged lepton data, the average of
JLAB resonance data, and neutrino data
at all $Q^2$. This model is currently used by current neutrino oscillation
experiments in the few GeV region.
\end{abstract}

\maketitle


%
The field of neutrino oscillation physics has progressed from
the discovery of neutrino oscillation~\cite{ATM}
to the era of precision measurements of  mass splitting and  mixing angles.
Currently, there are only poor measurements of
differential cross sections for neutrino interactions
in the few GeV region. This results in  large systematic uncertainties
in the extraction of mass splitting and  mixing parameters (e.g. by
the MINOS, NO$\nu$A , K2K and T2K experiments).
Therefore, reliable modeling of neutrino cross sections
at low energies is essential for precise (next generation) neutrino
oscillations experiments. In the few GeV region, there are three types
of neutrino interactions: quasi-elastic, resonance, and inelastic
scattering. It is very challenging to disentangle each contribution
separately, especially, resonance production versus deep inelastic
scattering (DIS) contributions. There are large non-perturbative QCD 
corrections
to the DIS contributions in this region.

Our approach is to relate neutrino interaction processes using
a quark-parton
model to precise charged-lepton scattering data.
In a previous communication~\cite{nuint02}, we showed that
our effective leading order model
using an improved scaling variable  $\xi_w$ describes
all deep inelastic scattering charged lepton-nucleon scattering data
including resonance data (SLAC/BCDMS/NMC/HERA/Jlab)~\cite{DIS,jlab}
from very high $Q^2$ to very low $Q^2$ (down to photo-production region),
as well as high energy  CCFR neutrino data~\cite{yangthesis}.

The proposed scaling variable, $\xi_w$
is derived using energy momentum conservation, assuming massless
initial state quarks bound in a proton of mass M.
\begin{eqnarray}
\label{eq:xi}
\xi_w &=& \frac{2x(Q^2+M_f{^2}+B)}
         {Q^{2} [1+\sqrt{1+(2Mx)^2/Q^2}]+2Ax},
\end{eqnarray}
here, $M_f$ is the final quark mass ( zero except for
charm-production  in
neutrino processes).
The parameter $A$ accounts for the higher order
(dynamic higher twist) QCD terms
in the form of an enhanced target mass term (the effects of the proton
target mass are already taken into account using the exact form
in the denominator of $\xi_w$ ). The parameter
$B$ accounts for the initial state quark transverse
momentum and final state quark  effective $\Delta M_f{^2}$
(originating from multi-gluon emission by quarks).
This parameter also allows us to describe the data also in
the photoproduction limit (all the way down to $Q^{2}$=0).

A brief summary of our effective leading order (LO) model is given as follows;
\begin{itemize}
  \item The GRV98 LO PDFs~\cite{grv98} are used to describe the $F_2$ 
data at high $Q^2$
	 region.
  \item  The scaling variable $x$ is replaced with the improved
         scaling variable $\xi_w$ (Eq.~\ref{eq:xi}).
\item  All PDFs are modified by $K$ factors to describe low $Q^2$
        data in the photoproduction limit.
  	\begin{eqnarray}
	\label{eq:kfac}
	 K_{sea}(Q^2) = \frac{Q^2}{Q^2 +C_s},~~~~~~~
	 K_{valence}(Q^2) =[1-G_D^2(Q^2)]
	            \left(\frac{Q^2+C_{v2}}
		      {Q^{2} +C_{v1}}\right),
	\end{eqnarray}
	 where $G_D$ = $1/(1+Q^2/0.71)^2$ is the  proton elastic form factor.
	At low $Q^2$, $[1-G_D^2(Q^2)]$
	is approximately $Q^2/(Q^2 +0.178)$. Different values of the 
$K$ factor
         are obtained for $u$ and $d$ quarks
  \item The evolution of the GRV98 PDFs is frozen at a
	value of $Q^2=0.80$ which is the minimum $Q^2$ value of this PDFs.
	Below this $Q^2$, $F_2$ is given by;
	\begin{eqnarray}
	     F_2(x,Q^2<0.8) = K(Q^2) \times F_2(\xi,Q^2=0.8)
	\end{eqnarray}

  \item Finally, using these effective GRV98 LO PDFs $(\xi_w)$
        we fit to all inelastic charged lepton
        scattering data (SLAC/BCDMS/NMC/H1) and photoproduction data
        on hydrogen and deuterium. Note that no resonance data is
        included in the fit.
         We obtain excellent fits with;
	 $A$=0.538, $B$=0.305, $C_{v1}^{d}$=0.202,
	$C_{v1}^{u}$=0.291, $C_{v2}^{d}$=0.255, $C_{v2}^{u}$=0.189,
	$C_{s1}^{d}$=0.621, $C_{s1}^{u}$=0.363,
	and $\chi^{2}/DOF=$1874/1574.
	Because of the $K$ factors to the PDFs, we find that
         the GRV98 PDFs need to be multiplied by  a factor of 1.015.

\end{itemize}

The measured structure functions data are corrected for
	the relative normalizations between the SLAC, BCDMS, NMC
	and H1 data. The deuterium data are corrected
	for nuclear binding effects~\cite{highx}.
	We also add a separate charm pair production contribution
	to lepton-nucleon scattering
	using the photon-gluon fusion
	model.
	This component is not
	necessary  at low energies, and is only
	needed
	to describe the highest $\nu$ HERA $F_2$ and photoproduction data
	(since the GRV98 LO PDFs do not include a charm sea).

\begin{figure}
\includegraphics[width=.47\columnwidth,height=.4\textheight]{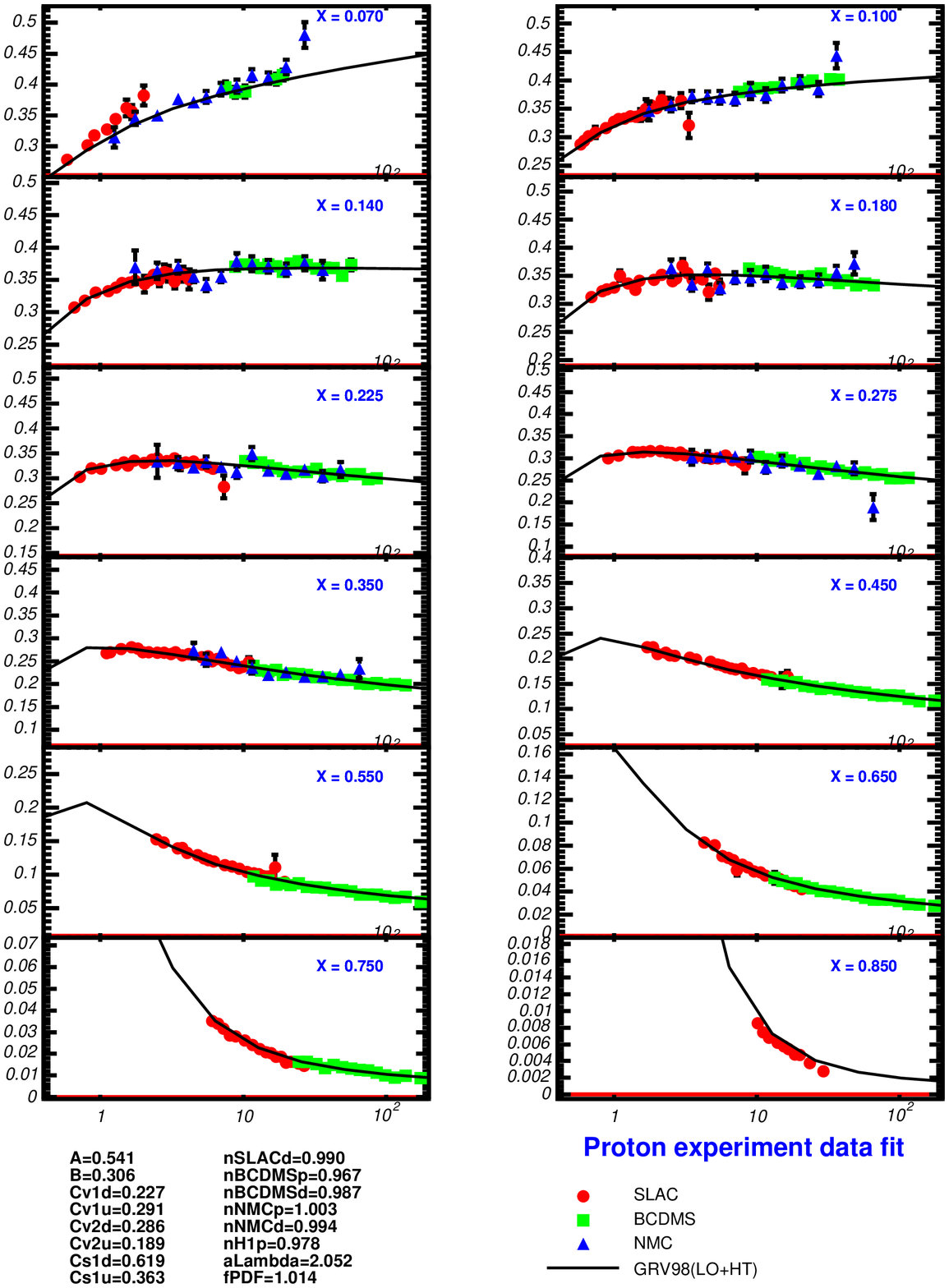}
\includegraphics[width=.47\columnwidth,height=.4\textheight]{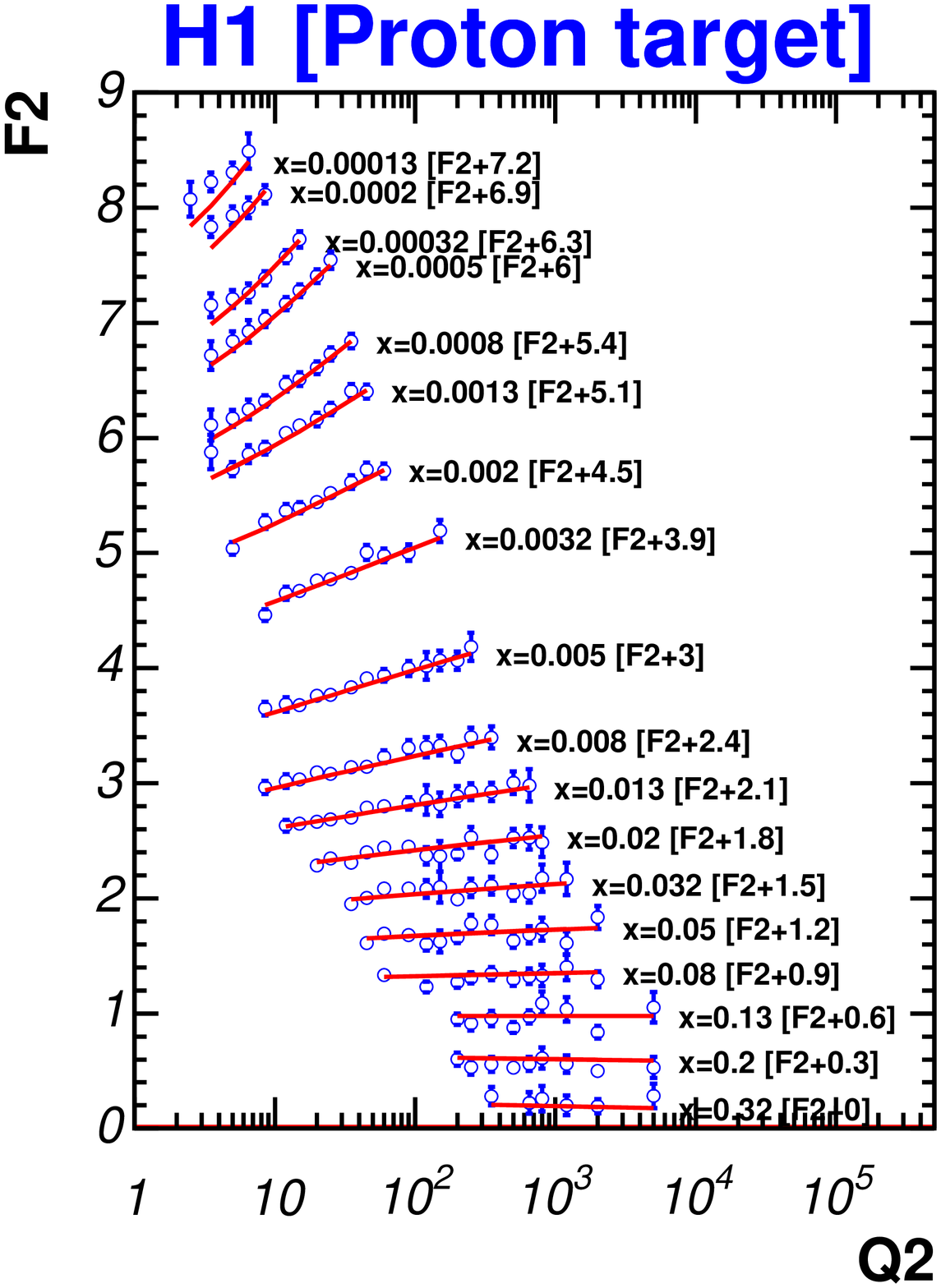}
\caption{ Comparisons of the predictions of our effective
LO model for $F_2$ to charged lepton inelastic scattering data.
[left] DIS $F_2$ proton data (SLAC, BCDMS, NMC), [right] H1 $F_2$ proton.}
\label{fig:f2fit}
\end{figure}

\begin{figure}
\includegraphics[width=.47\columnwidth,height=.4\textheight]{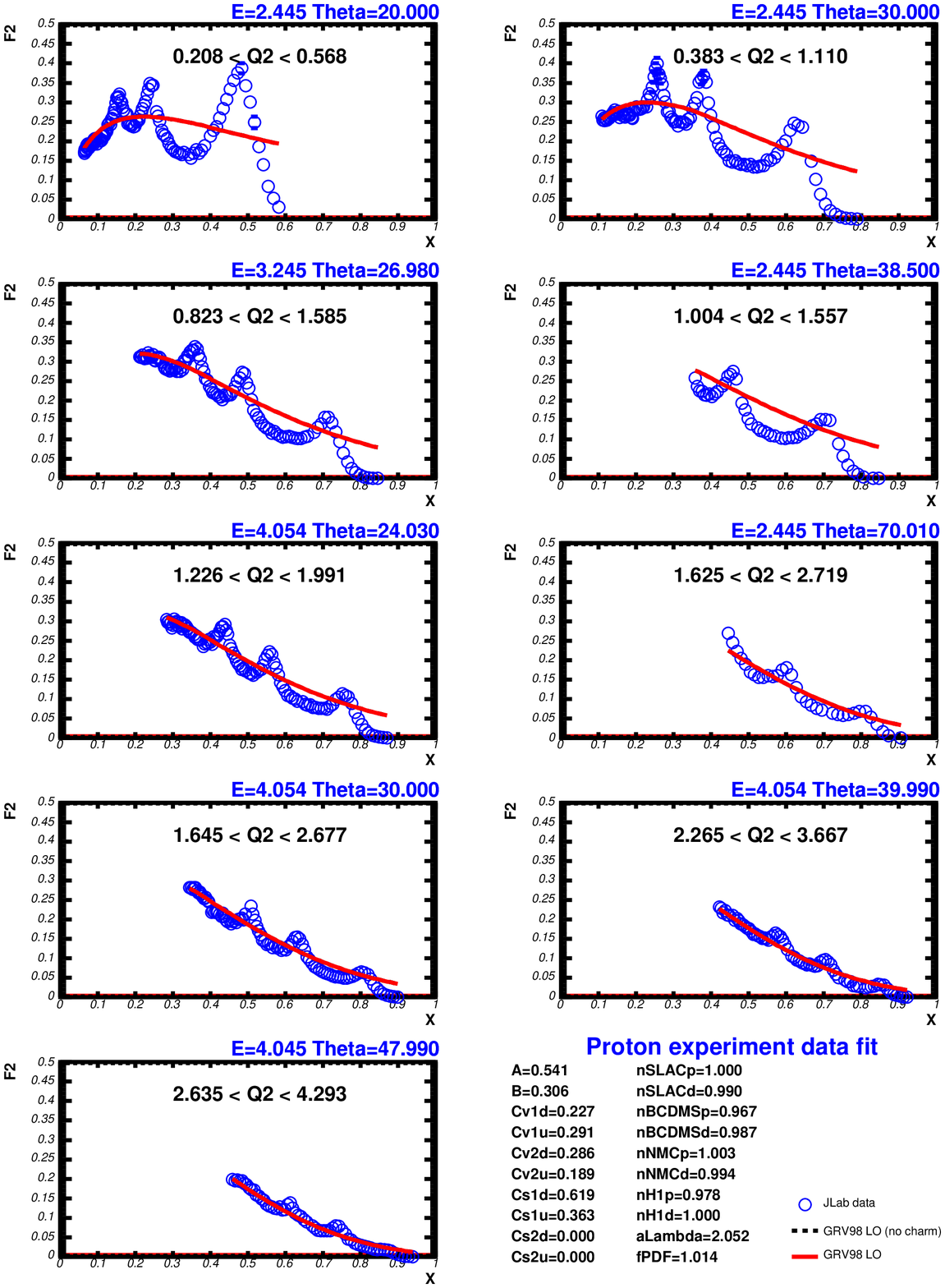}
\includegraphics[width=.47\columnwidth,height=.4\textheight]{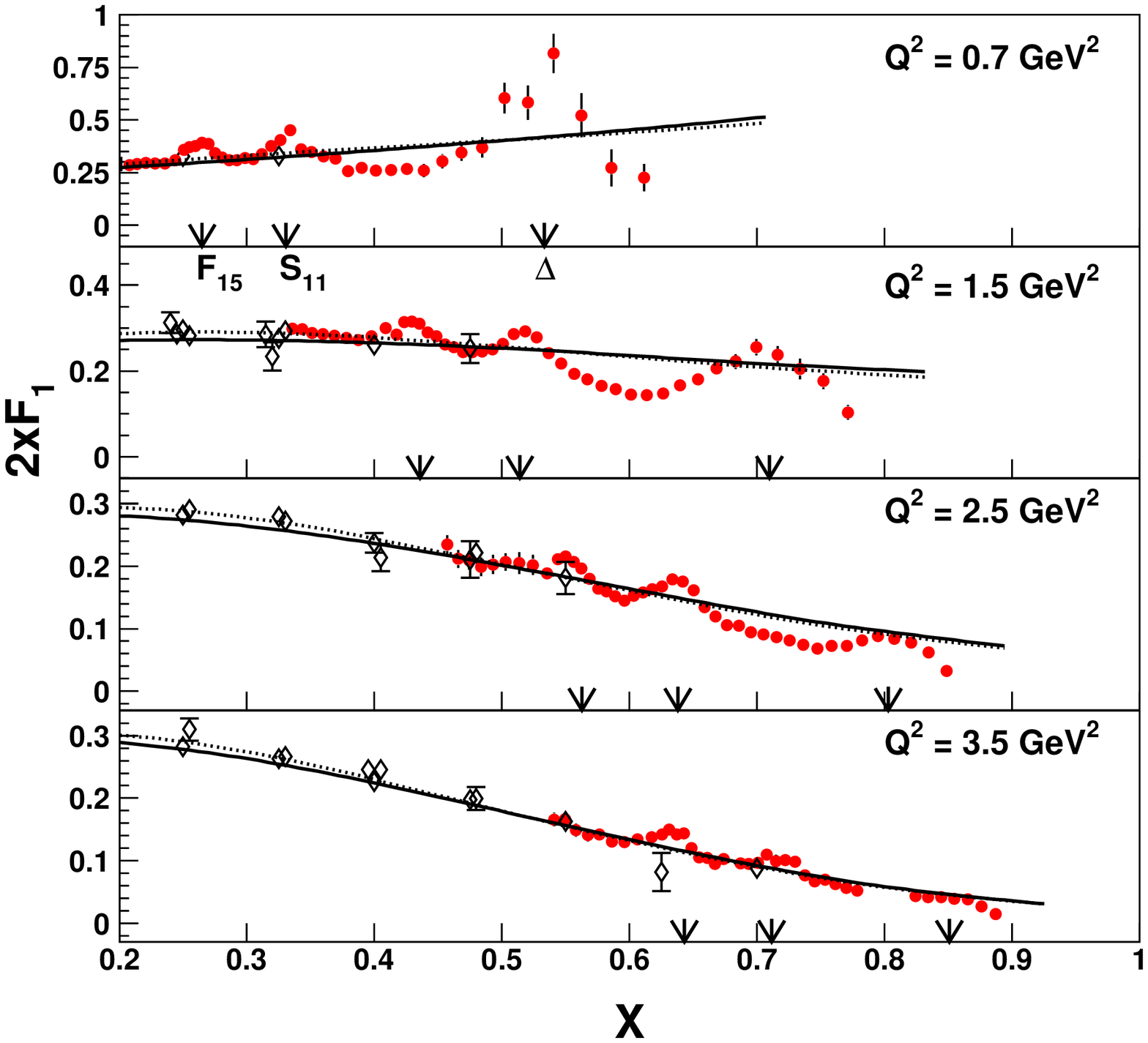}
\caption{ Comparisons of the predictions of our effective
LO model to resonance electro-production data on protons
(which was not included in our fit). Shown are
  $F_2$ proton [left], and $2xF_1$ proton data [right] (from the 
Jefferson Lab Hall C
E94-110 Collaboration).
The predictions for  $2xF_1$ are obtained from our model for $F_2$
with $R_{1998}$. In the right plot, the solid line uses GRV98 PDFs,
and the dashed line is our previous model using
GRV94 LO PDFs.}
\label{fig:predict}
\end{figure}

Our effective LO model
describes various DIS
(SLAC, BCDMS, NMC, and HERA) and photo-production data down to the
$Q^2=0$ limit.  Fig.~\ref{fig:f2fit} shows some of comparisons.
Furthermore, based on duality arguments~\cite{bloom},
it appears that this model also provides a reasonable description
of the average value of $F_2$ for SLAC and Jlab data in the resonance
region, as shown
in Fig.~\ref{fig:predict}[left]. Although no resonance data
has been included in our fit, our model gives a good
description of the most recent $2xF_1$ electron-proton Rosenbluth
separated  data in the resonance region from Jefferson Lab Hall C
E94-110 Collaboration~\cite{jlab_2xf1}, and also data from the first phase of
the JUPITER program at Jlab.
Our predictions for  $2xF_1$
are obtained  using our $F_2$ model and $R_{1998}$~\cite{R1998}.
For heavy targets, nuclear effects are important, especially at low $Q^2$.
Recent results from Jlab indicate that the Fe/D ratio in the
resonance region is the same as the Fe/D ratio
from DIS data for the same value of $\xi$ (or $\xi_w$).
Future Jlab experiments with deuterium
and heavy nuclear targets (e.g. JUPITER) will provide a high statistics data
in the resonance region which will be very important to improve our 
model at $Q^2$ region.

In neutrino scattering, in addition to the vector structure function,
there is an axial vector structure
function contribution.
At the $Q^2=0$ limit, the vector structure function
goes to zero, while the axial-vector part has a finite contribution.
At high $Q^2$, these two structure
functions are expected to be same. Thus, it is important
to understand the axial-vector contribution at low $Q^2$ by
comparing to future low energy neutrino data (e.g. MINER$\nu$A~\cite{minerva}).
As a preliminary step, we compare the CCFR and CDHSW~\cite{cdhsw} high energy
neutrino data with our model, assuming that the vector contribution
is the same as the axial vector contribution.
We find that the CCFR/CDHSW neutrino data are well described by our model.

We are currently working on constraining the low $Q^2$ axial vector
contribution using low energy CDHSW and CHORUS~\cite{chorus} data.
The form of the fits we plan to use is motivated
by the Adler sum rule~\cite{adler} for the axial vector contribution 
as follows:
\begin{eqnarray}
  K_{sea-ax}(Q^2) = \frac{Q^2+C_{2s-ax}}{Q^2 +C_{1s-ax}},~~~~~~~
  K_{valence}(Q^2) =[1-F_A^2(Q^2)]
        \left(\frac{Q^2+C_{2v-ax}} {Q^{2} +C_{1v-ax}}\right),
\end{eqnarray}
where $F_{A}(Q^2) = -1.267/(1+Q^2/1.00)^2$.
Nuclear effects for heavy target are also important and  may be different
for the vector and axial vector structure functions.
Future measurements
on the axial vector contribution from the  MINER$\nu$A 
experiment~\cite{minerva}
will be important in constraining this model.




\bibliographystyle{aipproc}   

\begin{thebibliography}{9}

\bibitem{ATM}  S. Fukuda {\em et al.}, Phys.
Rev. Lett. {\bf 85}, 3999 (2000); T. Toshito,   hep-ex/0105023.

\bibitem{nuint02} A. Bodek and U. K. Yang, hep-ex/0308007.

\bibitem{DIS} L. W. Whitlow {\it et al.}  (SLAC-MIT), Phys. Lett. B
{\bf 282}, 433 (1995);
A. C. Benvenuti {\it et al.} (BCDMS), Phys. Lett. B{\bf237}, 592 (1990);
  M. Arneodo {\it et al.}  (NMC), Nucl. Phys. B{\bf 483}, 3 (1997).
\bibitem{jlab}
C. Keppel, Proc. of the Workshop on Exclusive Processes
at High $P_T$, Newport News,
VA, May (2002).]

\bibitem{yangthesis}
U. K. Yang, Ph.D. thesis, Univ. of Rochester, UR-1583 (2001).

\bibitem{grv98} M. Gluck, E. Reya, A. Vogt, Eur. Phys. J {\bf C5}, 461 (1998).

\bibitem{highx} U. K. Yang and A. Bodek, Phys. Rev. Lett. {\bf 82}, 
2467 (1999),
  Eur. Phys. J. C{\bf 13}, 241 (2000).

\bibitem{bloom}E. D. Bloom and F. J. Gilman, Phys. Rev. Lett. {\bf
25}, 1140 (1970).

\bibitem{jlab_2xf1} Y. Liang {\em et al.}, nucl-ex/0410027.

\bibitem{R1998}
K. Abe {\em et al.}, {\it Phys. Lett.} {\bf B452}, 194 (1999).
%
\bibitem{cdhsw}
P. Berge {\em et al.} (CDHSW), Zeit. Phys. {\bf C49}, 607 (1991).

\bibitem{chorus} R. Oldeman, Proc. of
30th International Conference on High-Energy Physics (ICHEP 2000),
Osaka, Japan, 2000.

\bibitem{adler} S. Adler, Phys. Rev. {\bf 143}, 1144 (1966);
F. Gillman, Phys. Rev. {\bf 167}, 1365 (1968).
%
\bibitem{minerva} MINER$\nu$A Proposal, D. Drakoulakos {\em et al.}, 
hep-ex/0405002

\end{thebibliography}

\end{document}